\documentclass[showpacs,preprintnumbers,10pt,twocolumn]{revtex4}
\usepackage{dcolumn}
\usepackage{bm}
\usepackage{graphicx}
\usepackage{amsfonts}
\usepackage{amssymb}
\usepackage{amsmath}

\begin{document}

\title{Deterministically entangling distant nitrogen-vacancy centers by a \\
nanomechanical cantilever}
\author{Z. Y. Xu$^{1,2}$}
\author{Y. M. Hu$^{1,2}$}
\author{W. L. Yang$^{1,2}$}
\author{M. Feng$^{1}$}
\email{mangfeng@wipm.ac.cn}
\author{J. F. Du$^{3}$}
\email{djf@ustc.edu.cn}
\affiliation{$^{1}$State Key Laboratory of Magnetic Resonance and Atomic and Molecular
Physics, Wuhan Institute of Physics and Mathematics, Chinese Academy of
Sciences, Wuhan 430071, China}
\affiliation{$^{2}$Graduate School of the Chinese Academy of Sciences, Beijing 100049,
China}
\affiliation{$^{3}$Hefei National Laboratory for Physics Sciences at Microscale and
Department of Modern Physics, University of Science and Technology of China,
Hefei, 230026, China}

\begin{abstract}
We present a practical scheme by global addressing to
deterministically entangle negatively charged nitrogen-vacancy (N-V)
centers in distant diamonds using a nano-mechanical cantilever with
the magnetic tips strongly coupled to the N-V electron spins.
Symmetric Dicke states are generated as an example, and the
experimental feasibility and challenge of our scheme are discussed.
\end{abstract}

\maketitle

Over past years, negatively charged nitrogen-vacancy (N-V) centers in
diamond have been considered as one of the most promising candidates for
solid-state quantum information processing (QIP) due to their long coherence
time at room temperature \cite{coherence-room-teper}. The intensive
experimental studies have been paid on quantum logical gating \cite{CNOT},
state storage and transfer \cite{register}, and preparation of entangled
states \cite{GHZ}, using the electron-spin and the nearby nuclear spins
within a single diamond.

In the context of a large-scale diamond-based QIP, the generation of
multipartite entangled states among distant N-V centers is of particular
importance and experimental challenge. One of the proposed approaches is to
use the measurement-based QIP scheme \cite{measure-based}. By detecting the
emitted photons which have been entangled with different N-V electron spins,
distant N-V centers can be entangled in a nondeterministic way. Such an
approach has the major advantage that no direct interactions among qubits
are required, which implies a relatively easy control. An alternative is the
scalable optical coupling between N-V centers and high-Q microcavities \cite%
{micro-cavity}. By sharing the common modes of a microcavity or sharing
flying single-photon pulses, the remote N-V centers could be coherently
manipulated or entangled, which may open up the possibilities of N-V-based
scalable QIP.

\begin{figure}[tbp]
\centering
\includegraphics[width=3.3in]{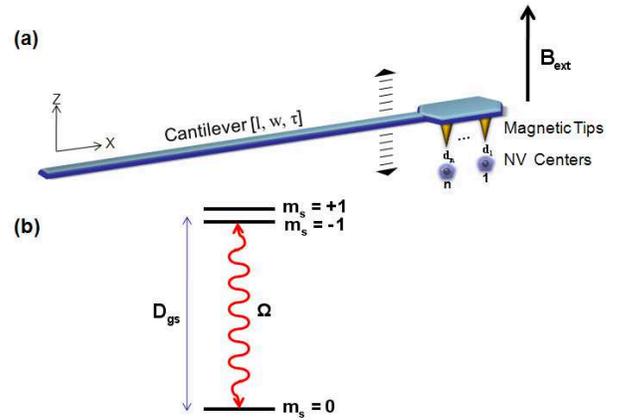}
\caption{(Color online) (a) Schematic of the physical system
consisting of an array of N-V centers strongly coupled to the
quantized motion of a cantilever with length $l$, width $w$ and
thickness $\protect\tau $. An external magnetic field
$\mathbf{B}_{ext}$ is applied along the symmetry axis of the N-V
centers. (b) The energy level diagram for the ground triplet
state of the electron spin within a N-V center, with a zero-field splitting D$%
_{gs}$ between spin levels $m_{s}=0$ and $m_{s}=\pm 1.$ We restrict our
study to the down state $m_{s}=0 $ and the up state $m_{s}=-1$ by driving a
microwave field.}
\label{fig1}
\end{figure}

To have an efficient processing of quantum information, however, we prefer
deterministic manipulations. In the present work, we will show the
possibility to entangle multiple N-V centers by a nano-mechanical cantilever
in a deterministic fashion. We had noticed a recent work for a strong
magnetic coupling between electron-spin degrees of freedom of a N-V center
and a nano-mechanical resonator \cite{magnetic-tip, cooling}. In our case,
we have a line of N-V centers, each of which is located in a nanoscale
diamond. We will show how to generate the symmetric Dicke states \cite%
{Dicke-state}, denoted by $|D_{n}^{(m)}\rangle $ with equally weighted
superposition of all permutations of \textit{m} excitations among \textit{n}
particles, by global addressing. Such multipartite entangled states own
great properties of robustness against decoherence and particle loss and
have been observed experimentally with linear optics systems \cite%
{Dicke-experiment}. The favorable features of our scheme include the
deterministic preparation of entangled electron-spin states between N-V
centers and the potential to be scaled up with near future technologies.
Besides, as a scheme with global addressing, only a microwave field is
required during the entanglement preparation, which greatly reduces the
experimental time and difficulty. Furthermore, our proposed scheme could be
realized with very high fidelity due to our fast manipulations and the long
decoherence time of electron spins in N-V centers.

As sketched in Fig. 1(a), an array of evenly spaced magnetic tips are
attached at the end of a nano-mechanical cantilever, under which are \textit{%
n} one-to-one correspondent N-V centers. In order to entangle the electron
spins in spatially separate N-V centers by a cantilever, we have to couple
the electron spins to the motion of the resonator. This can be achieved by
exposing the electron spins to the magnetic field gradient created by the
magnetic tips. For clarity of our description, we label $\lambda _{j}=\chi
_{e}B_{g}^{j}z_{0}^{j}$ $(j=1,2,\cdots n)$ as the coupling strength, where $%
B_{g}^{j}=|\partial _{z}\mathbf{B}_{tip_{j}}|$, $z_{0}^{j}$ is the
zero-point amplitude of the cantilever regarding the \textit{j}th tip and $%
\chi _{e}$ is the charge-mass ratio of an electron. For simplicity, we
assume that all the\textit{\ }N-V centers are equally coupled to the
cantilever ($\lambda _{j}=\lambda $), which could be achieved by elaborately
adjusting the magnetic field gradient of the tips and the distances between
the tips and the corresponding N-V centers. Furthermore, in most part of the
paper, we will neglect the influence from the nuclear spins, which implies a
pure N-V center without $^{13}C$ doped or with well polarized nuclear spins
of $^{13}C$ or $^{15}N$ \cite{polarization}. For our purpose, we encode
qubits in two sublevels: $m_{s}=0$ and $m_{s}=-1$ of the ground triplet
state, and a microwave field will be employed to drive the two sublevels. As
a result, the Hamiltonian of the whole system can be written in units of $%
\hbar =1$ as
\begin{equation}
\begin{array}{ll}
H= & \sum_{j=1}^{n}\omega _{0}|-1\rangle _{j}\left\langle -1\right\vert +\nu
a^{\dag }a \\
& +\sum_{j=1}^{n}\frac{\Omega _{j}(t)}{2}\left( |0\rangle _{j}\left\langle
-1\right\vert e^{i\omega _{j}t}+|-1\rangle _{j}\left\langle 0\right\vert
e^{-i\omega _{j}t}\right)  \\
& +\sum_{j=1}^{n}\frac{\lambda }{2}(a+a^{\dag })\sigma _{z}^{j},%
\end{array}
\label{1}
\end{equation}%
where $\omega _{0}\approx D_{gs}-\chi _{e}|\mathbf{B}_{ext}|$ is the energy
splitting between sublevels $|0\rangle $ and $|-1\rangle $ under the
externally applied magnetic field $\mathbf{B}_{ext}$ along z axis, i.e., the
quantization axis, $D_{gs}$ (around 2.87 GHz \cite{Dgs}) is the zero-field
splitting between the spin levels $m_{s}=0$ and $m_{s}=\pm 1$ shown in Fig.
1(b). $\nu $ and $a$ $\left( a^{\dag }\right) $ are the frequency and
annihilation (creation) operator of the motional mode of the cantilever.$\
\Omega _{j}(t)$ and $\omega _{j}$ are the Rabi frequency and the frequency
of the microwave regarding the cantilever coupling to the $j$th N-V center,
and $\sigma _{z}^{j}=|-1\rangle _{j}\left\langle -1\right\vert -$ $|0\rangle
_{j}\left\langle 0\right\vert $. For simplicity, we may assume in following
treatment $\Omega _{j}(t)=\Omega (t)$ and $\omega _{j}=\omega $ $%
(j=1,2,\cdots n)$. So the Hamiltonian of the total system in the interaction
picture is
\begin{equation}
\begin{array}{ll}
H_{I}= & \sum_{j=1}^{n}\frac{\Omega (t)}{2}[|0\rangle _{j}\left\langle
-1\right\vert e^{i\left( \omega -\omega _{0}\right) t} \\
& +|-1\rangle _{j}\left\langle 0\right\vert e^{-i\left( \omega -\omega
_{0}\right) t}] \\
& +\sum_{j=1}^{n}\frac{\lambda }{2}(ae^{-i\nu t}+a^{\dag }e^{i\nu t})\sigma
_{z}^{j}.%
\end{array}
\label{2}
\end{equation}%
If $\omega =\omega _{0},$ Eq. (2) in a new basis $|\pm \rangle =\left(
|-1\rangle \pm |0\rangle \right) /\sqrt{2}$ reads as $\sum\nolimits_{j=1}^{n}%
\frac{\Omega (t)}{2}\tilde{\sigma}_{z}^{j}+\sum\nolimits_{j=1}^{n}\frac{%
\lambda }{2}(ae^{-i\nu t}+a^{\dag }e^{i\nu t})\tilde{\sigma}_{x}^{j},$ where
$\tilde{\sigma}_{x}^{j}=|+\rangle _{j}\left\langle -\right\vert +$ $%
|-\rangle _{j}\left\langle +\right\vert $ and $\tilde{\sigma}%
_{z}^{j}=|+\rangle _{j}\left\langle +\right\vert -$ $|-\rangle
_{j}\left\langle -\right\vert .$ Considering $\Omega (t)$ is comparable to $%
\nu ,$ we rewrite Eq. (2) in a rotating frame under rotating wave
approximation, obtaining%
\begin{equation}
\tilde{H}_{I}=\sum\nolimits_{j=1}^{n}\frac{\lambda }{2}\left\{ a|+\rangle
_{j}\left\langle -\right\vert \exp \left[ -i\int_{t_{i}}^{t}\delta (t)dt%
\right] +H.c.\right\} ,  \label{3}
\end{equation}%
with $\delta (t)=\nu -\Omega (t)$.

From the viewpoint of QIP, the vibrational mode of a cantilever could be
considered as a common data bus for distant N-V centers. Therefore, using
Eq. (3), we may achieve various kinds of multiple entangled states. As an
example, we will show below how to prepare symmetric Dicke states of the
electron spins in distant N-V centers by global microwave addressing.

We consider a Si cantilever beam of length $l=5$ $\mu $m$,$ width $w=50$ nm
and thickness $\tau =50$ nm with effective mass $M\simeq 7.28\times 10^{-18}$
kg, resonant frequency of the lowest-order flexural mode $\nu \simeq 14.87$
MHz and the zero-point amplitude of fluctuation $\sqrt{1/(2M\nu )}\simeq
6.98\times 10^{-13}$ m \cite{cantilever}. The N-V-cantilever coupling
strength $\lambda \simeq 0.96$ MHz has been achieved under a magnetic field
gradient $7.8\times 10^{6}$ Tm$^{-1}$ at a distance $\simeq 25$ nm away from
the tip \cite{magnetic-tip,Gradient}$.$ In addition, the distance between
the nearest-neighbor tips is set to be around 150 nm.

We will generate two nontrivial three-particle symmetric Dicke states: $%
|D_{3}^{(1)}\rangle=$ $\left( |+--\rangle+|-+-\rangle+|--+\rangle \right) /
\sqrt{3}$ and $|D_{3}^{(2)}\rangle=$ $\left( |++-\rangle
+|+-+\rangle+|-++\rangle\right) /\sqrt{3}$. We get started from Eq. (3),
which can be rewritten as $\tilde{H}_{I}=\frac{\lambda }{2}%
\sum\nolimits_{j=1}^{3}(a|+\rangle_{j}\left\langle -\right\vert +a^{\dag
}|-\rangle_{j}\left\langle +\right\vert
)-\delta(t)\sum\nolimits_{j=1}^{3}|+\rangle_{j}\left\langle +\right\vert$
after performing a time-dependent phase transformation.

\begin{figure}[htbp]
\centering
\includegraphics[width=3.5in]{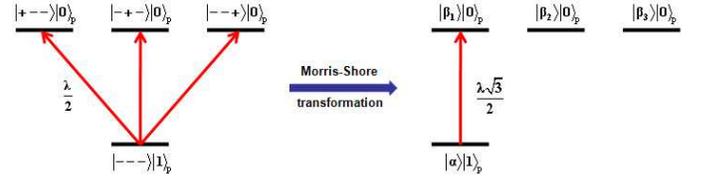}
\caption{(Color online) Left: The energy level pattern of a string of three
NV centers with an initial state $|---\rangle |1\rangle _{p}$ globally
driven by a microwave field, where we set $\protect\delta (t)=0$ and $%
|\rangle _{p}$ is for the motional state of the cantilever. Right: The MS
transformation produces a two-level subsystem where $|\protect\beta %
_{1}\rangle =$ $|D_{3}^{(1)}\rangle $, $|\protect\alpha \rangle =|---\rangle
$ and two decoupled dark states $|\protect\beta _{2}\rangle =(|+--\rangle
-2|-+-\rangle +|--+\rangle )/\protect\sqrt{6}$ and $|\protect\beta %
_{3}\rangle =(|+--\rangle -|--+\rangle )/\protect\sqrt{2}$.}
\label{fig2}
\end{figure}

Let us first consider how to prepare $|D_{3}^{(1)}\rangle $. We assume that
the vibrational state of the cantilever has been cooled to the Fock state $%
|1\rangle _{p}$ \cite{magnetic-tip,cooling} and the electron-spin states of
N-V centers are initially prepared in $|---\rangle .$ By means of
Morris-Shore (MS) transformation \cite{MS}, it is possible to factorize the
Hamiltonian into a smaller closed subspace \cite{Dicke-collective}. As shown
in Fig. 2, in MS basis the original system is decomposed to a two-level
subsystem plus two independent single-level subsystems. The effective
Hamiltonian of the two-level subsystem is written as
\begin{equation}
H_{eff}^{(1)}=\left(
\begin{array}{cc}
-\delta (t) & \frac{\sqrt{3}}{2}\lambda  \\
\frac{\sqrt{3}}{2}\lambda  & 0%
\end{array}%
\right)   \label{4}
\end{equation}%
in the basis \{$|\beta _{1}\rangle |0\rangle _{p},|\alpha \rangle |1\rangle
_{p}$\}, where $|\beta _{1}\rangle =$ $|D_{3}^{(1)}\rangle $ and $|\alpha
\rangle =|---\rangle .$ The top state $|\beta _{1}\rangle $ is connected to
the initial state $|\alpha \rangle $ with an effective coupling strength $%
\sqrt{3}\lambda /2$ and can be interchanged by a microwave field addressing
the three N-V centers globally. When $\delta (t)$ is changed to zero (by
using a nearly square pulse of the microwave), the desired $%
|D_{3}^{(1)}\rangle $ can be realized at t=$\pi /(\sqrt{3}\lambda )$ with
unit probability.

\begin{figure}[tbph]
\centering
\includegraphics[width=3.5in]{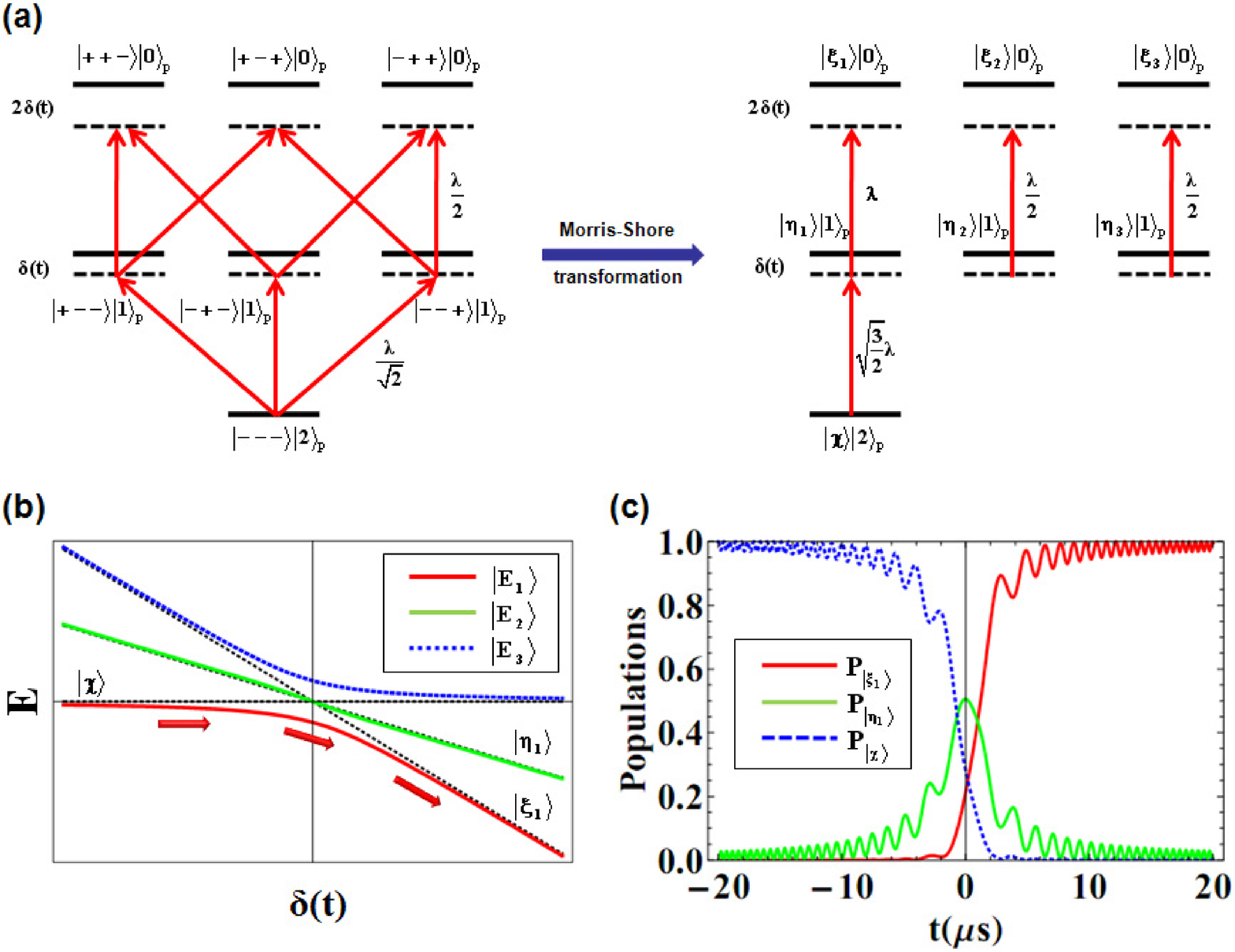}
\caption{(Color online) (a) Left: The energy level pattern of a string of
three NV centers with an initial state $|---\rangle |2\rangle _{p}$ globally
driven by a microwave field. Right: The MS transformation produces a
three-level subsystem where $|\protect\xi _{1}\rangle =|D_{3}^{(2)}\rangle ,$
$|\protect\eta _{1}\rangle =|D_{3}^{(1)}\rangle $, $|\protect\chi \rangle
=|---\rangle $ and two decoupled two-level subsystems $|\protect\xi %
_{2}\rangle =(|++-\rangle -2|+-+\rangle +|-++\rangle )/\protect\sqrt{6},$ $|%
\protect\eta _{2}\rangle =(|+--\rangle -2|-+-\rangle +|--+\rangle )/\protect%
\sqrt{6}$ and $|\protect\xi _{3}\rangle =(|++-\rangle -|-++\rangle )/\protect%
\sqrt{2},$ $|\protect\eta _{3}\rangle =(|+--\rangle -|--+\rangle )/\protect%
\sqrt{2}$. (b) Adiabatic evolution of the three-level subsystem, where the
three black-dotted lines represent $|\protect\xi _{1}\rangle ,$ $|\protect%
\eta _{1}\rangle $ and $|\protect\chi \rangle $ respectively. The up state $|%
\protect\xi _{1}\rangle $ and ground state $|\protect\chi \rangle $
can be interchanged by adiabatic passages (arrows) under the
adiabatic
condition. (c) Populations of states $|\protect\xi _{1}\rangle ,$ $|\protect%
\eta _{1}\rangle ,$ $|\protect\chi \rangle $ during the adiabatic evolution.}
\label{fig3}
\end{figure}

Now we turn to the preparation of $|D_{3}^{(2)}\rangle .$ In contrast to $%
|D_{3}^{(1)}\rangle $ preparation, the state of the whole system in this
case is initially prepared in $|---\rangle |2\rangle _{p}.$ After performing
MS transformation, we obtain a three-level subsystem plus two additional
two-level subsystems [See Fig. 3(a)]. The effective Hamiltonian reads as
\begin{equation}
H_{eff}^{(2)}=\left(
\begin{array}{ccc}
-2\delta (t) & \lambda  & 0 \\
\lambda  & -\delta (t) & \lambda \sqrt{\frac{3}{2}} \\
0 & \lambda \sqrt{\frac{3}{2}} & 0%
\end{array}%
\right)   \label{5}
\end{equation}%
in the basis $\left\{ |\xi _{1}\rangle |0\rangle _{p},|\eta _{1}\rangle
|1\rangle _{p},|\chi \rangle |2\rangle _{p}\right\} ,$ where $|\xi
_{1}\rangle =|D_{3}^{(2)}\rangle ,$ $|\eta _{1}\rangle =|D_{3}^{(1)}\rangle $
and $|\chi \rangle =|---\rangle $. It is easy to check by setting $\delta
(t)=0$ that the population probability of $|D_{3}^{(2)}\rangle $ will only
reach 0.95. In order to generate $|D_{3}^{(2)}\rangle $ with higher
probability, we will employ adiabatic techniques as below. The top state $%
|\xi _{1}\rangle $ and the bottom state $|\chi \rangle $ in Fig. 3(b) can be
interchanged through adiabatic passages under the adiabatic condition: $%
\underset{t\in \lbrack t_{i},t_{f}]}{\max }\left\vert \left\langle
E_{l}(t)\right\vert \partial _{t}H_{eff}^{(2)}|E_{k}(t)\rangle
/[E_{k}(t)-E_{l}(t)]^{2}\right\vert \ll 1$ ($l\neq k$)$,$ where $%
|E_{k}\rangle $ and $E_{k}$ are the instantaneous eigenstates and
eigeN-Values of Eq. (5). In Fig. 3(c), the populations of $|\xi _{1}\rangle ,
$ $|\eta _{1}\rangle $ and $|\chi \rangle $ are numerically calculated from
Eq. (5), where we have assumed a purely linear microwave chirp (using
sawtooth pulse of the microwave). By addressing the three N-V centers
globally, we could achieve the population probability of $%
|D_{3}^{(2)}\rangle $ to be higher than 0.99 through this adiabatic passage.

With similar steps to the above cases for three N-V centers, a
generalization to $|D_{n}^{(m)}\rangle $ with $n>m>$ 3 is straightforward.

In the above treatment, we have assumed the N-V centers to be equally
coupled to the cantilever. In realistic case, however, the coupling should
be slightly different due to imprecision. Figure 4(a) demonstrates a
simplified case for the fidelities of the produced states under the relative
error of the coupling strength $\Delta $, defined by $\lambda
_{n}=[1-(n-1)\Delta ]\lambda ,$ where the fidelities still remain high in
the case of the small relative error.

The main decoherence sources associated with our implementation are
dephasing of the electron spins and the intrinsic heating of the cantilever.
Since our operation for the symmetric Dicke states could be accomplished on
the time scale of $1.8\sim 30$ $\mu $s, which is much shorter than the
reported electron-spin relaxation time $T_{1}=6$ ms \cite{GHZ} as well as
the dephasing time $T_{2}=350$ $\mu s$ \cite{T2} induced by the nuclear-spin
fluctuation inside a N-V center. Thus, the influence on the N-V
electron-spin from the intrinsic damping and dephasing is negligible. On the
other hand, as the cantilever is ultra-sensitive to the magnetic force
generated between the tips and the N-V electron spins \cite%
{cantilever,Gradient,sensitive}, we will concentrate on the heating effect
of the cantilever by the master equation $\partial _{t}\rho
=-i[H_{eff}^{(k)},\rho ]+$\L $\rho $ with $k=$1, 2 [corresponding to Eqs.
(4) and (5), respectively], where \L $\rho =\Gamma (\bar{n}_{th}+1)\left(
2a\rho a^{\dag }-a^{\dag }a\rho -\rho a^{\dag }a\right) /2+\Gamma \bar{n}%
_{th}\left( 2a^{\dag }\rho a-aa^{\dag }\rho -\rho aa^{\dag }\right) /2,$ $%
\bar{n}_{th}=1/[\exp (\frac{\hbar \nu }{k_{B}T})-1]$ and $k_{B}$ is the
Boltzmann constant. In general, the heating factor of the cantilever is
described by $\Gamma =\nu /Q$ \cite{cantilever}$,$ where $\nu $ and $Q$ are
the vibrational frequency and the quality factor of a cantilever. Numerical
calculations are performed for the fidelities of the Dicke states in Fig.
4(b), where the fidelities of $|D_{3}^{(1)}\rangle $ and $%
|D_{3}^{(2)}\rangle $ could reach 0.998 and 0.996 respectively in the case
of the $Q$ value $10^{5}$. If the $Q$ value is lowered to 10$^{4}$, the
fidelity drops to 0.985 for $|D_{3}^{(1)}\rangle $ and to 0.957 for $%
|D_{3}^{(2)}\rangle $.

\begin{figure*}[tbp]
\centering
\includegraphics[width=6in]{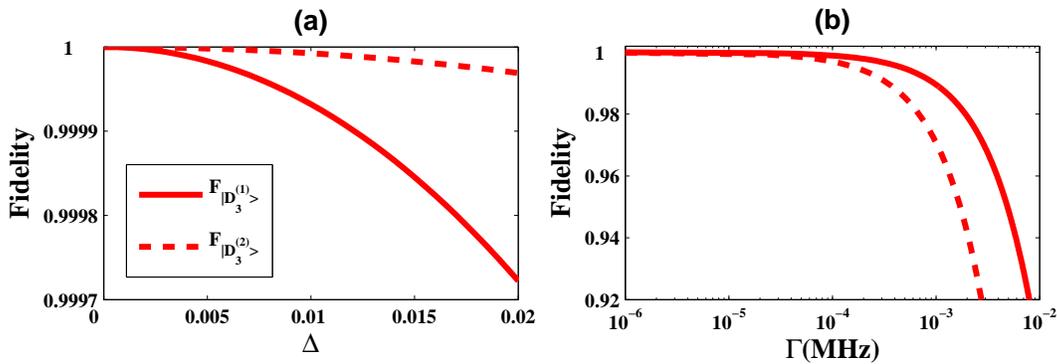}
\caption{(Color online) The fidelity of symmetric Dicke states $%
|D_{3}^{(1)}\rangle $ (solid line) and $|D_{3}^{(2)}\rangle $ (dashed line)
versus (a) relative error $\Delta $ of the coupling strength; (b) heating
factor $\Gamma $ of the cantilever.}
\label{fig4}
\end{figure*}

We address some remarks for experimental issues of our scheme. To make sure
that the N-V centers are all strongly and nearly equally coupled to the
cantilever, we have to restrict the tips within a small region near the head
of the cantilever, which implies a limitation of the number for available
N-V centers (at most tens of N-V centers in our present design). So a
potential for scalability with our scheme might be to combine with the
non-deterministic method \cite{measure-based}. Besides, we require nanoscale
diamonds with a single N-V center in each and the N-V centers should be
located very close to the surface of the diamonds. To our knowledge, such
diamond nanoparticles are not in principle unavailable although they have
not yet been achieved experimentally. On the other hand, due to the small
spacing between adjacent N-V centers, individual addressing with a microwave
field is of difficulty. As a result, although entanglement could be achieved
with global operations, to carry out universal quantum computing, we may
have to employ the near-field optical technique \cite{near} with spatial
resolution of the order of tens of nanometers, which had been widely used in
optics and surface science, and recently also employed for QIP \cite{feng}.
Alternatively, a magnetic field gradient along x-axis would help to
distinguish different qubits in frequency space. Thus universal QIP is
available, and robust quantum gating, such as composite pulse method \cite%
{jones} or bang-bang control \cite{viale}, employed previously could be
straightforwardly applied to our scheme. Finally, cooling a cantilever with
many magnetic tips to a required low temperature (114 $\mu K$ in our design)
could be of great technical difficulty with current technology but may be
resolved by recently presented cooling schemes \cite{magnetic-tip,cooling}.

Further consideration would iN-Volve nuclear spins which are good at storing
quantum information due to the long decoherence time \cite{nuclear-coherence}%
. In this sense, we may also consider to encode the qubits in the nuclear
spins, and employ the electron spins as ancillas. The hyperfine interaction
helps to transfer the generated entanglement of electron spins to the
corresponding nuclear spins of N-V centers \cite{register}.

To conclude, we have proposed a scheme to deterministically entangle a
string of N-V centers by strongly coupling to the quantized motion of a
nano-mechanical cantilever. Due to the long relaxation and dephasing time of
the electron spins in N-V centers together with the weak intrinsic heating
effects of the cantilever, our present scheme would enable a promising way
for future N-V-center-based QIP.

The work is supported by NNSFC under Grant No. 10774042, by CAS, and by
NFRPC.

\end{document}